\documentclass[superscriptaddress,amsmath,amssymb,aps,prl,titlepage,reprint]{revtex4-1}%
\usepackage[utf8]{inputenc}
\usepackage{amsfonts}
\usepackage{graphicx}
\usepackage{float}
\usepackage{subfigure}
\usepackage{natbib}
\usepackage{lipsum}

\usepackage{color,soul} 

\begin{document}

\title{Spatial Mapping of Torques within a Spin Hall Nano-oscillator}
\author{T. M. Spicer}
\author{P. S. Keatley}
\author{T. H. J. Loughran}
\affiliation{Department of Physics and Astronomy, University of Exeter, EX4 4QL,United Kingdom}

\author{M. Dvornik}
\author{A. A. Awad}
\author{P. D\"{u}rrenfeld}
\author{A. Houshang}
\author{M. Ranjbar}
\affiliation{Department of Physics, University of Gothenburg, 412 96 Gothenburg, Sweden}

\author{J. \r{A}kerman}
\affiliation{Department of Physics, University of Gothenburg, 412 96 Gothenburg, Sweden}
\affiliation{Materials Physics, School of ICT, KTH-Royal Institute of Technology, Electrum 229, 164 40 Kista, Sweden}

\author{V. V. Kruglyak}
\author{R. J. Hicken}
\affiliation{Department of Physics and Astronomy, University of Exeter, EX4 4QL,United Kingdom}

\begin{abstract}
Time-resolved scanning Kerr microscopy (TRSKM) was used to study precessional magnetization dynamics induced by a radio frequency (RF) current within a Al$_2$O$_3$/Py(5 nm)/Pt(6 nm)/Au(150 nm) spin Hall nano-oscillator structure. The Au layer was formed into two needle-shaped electrical contacts that concentrated the current in the centre of a Py/Pt mesa of 4 $\mu$m diameter. Due to the spin Hall effect, current within the Pt layer drives a spin current into the Py layer, exerting a spin transfer torque (STT). By injecting RF current, and exploiting the phase-sensitivity of TRSKM and the symmetry of the device structure, the STT and Oersted field torques have been separated and spatially mapped.  The STT and torque due to the in-plane Oersted field are observed to exhibit minima at the device centre that is ascribed to spreading of RF current that is not observed for DC current.  Torques associated with the RF current may destabilise the position of the self-localised bullet mode excited by the DC current, and inhibit injection locking. The present study demonstrates the need to characterise both DC and RF current distributions carefully.
\end{abstract}

\maketitle

Spin torque oscillators (STOs)\cite{Chen2016} are nanoscale magnetoresistive devices of great promise for use in microwave assisted magnetic recording \cite{Shiroishi2009}, microwave frequency telecommunications \cite{Karenowska2015}, and neuromorphic computing \cite{Grollier2016}.  Injection of DC current $I_{DC}$ generates spin transfer torque (STT) that excites precession of the constituent magnetic moments. The magnetoresistance (MR)  leads to an RF voltage across the device. Within spin Hall nano-oscillator (SHNO) devices, a charge current is converted into a pure spin current, via the spin Hall effect (SHE) \cite{Kato2004,Hirsch1999,Jungwirth2012,Hoffmann2013}, which exerts a STT upon the active magnetic layer. Decoupling of spin and charge currents provides additional freedom in the device design, the magnetic materials used  \cite{Nouriddine2011,Durrenfeld2015,Mazraati2016,Zahedinejad2018} (including electrical insulators \cite{Hamadeh2014}) and the precessional modes excited \cite{Dvornik2018}.  If charge current flows parallel to the plane, without a top contact obscuring the active region, then optical techniques can probe the magnetization dynamics \cite{Demidov2012,Awad2016}. However the spatial current distribution may be highly non-uniform, and thermal effects may modify both the current distribution and the torques \cite{Liu2013,Demidov2014,Ulrichs2014}. Knowledge of the Oersted torque and STT is critical for understanding how auto-oscillations may be excited, or locked to a reference signal, and until now it has been impossible to probe their spatial distribution directly.

STT-ferromagnetic resonance (STT-FMR) is widely used to characterise spintronic devices. Radio frequency (RF) current $I_{RF}$ is injected to excite the magnetisation, and mixes with the oscillatory MR response to generate a DC mixing voltage $V_{mix}$, which is recorded as the applied magnetic field or the frequency of $I_{RF}$ is varied. Analysis of the resonance field, or frequency, and linewidth allows the torques acting upon the magnetization to be determined \cite{Skinner2013,Tulapurkar2005,Kalarickal2006,Sankey2007,Kubota2008,Liu2011,Brataas2012,Dumas2014}. However, $V_{mix}$ may vanish for certain magnetic field configurations, and represents a spatial average of magnetization dynamics that may be highly inhomogeneous. 

Recently, time resolved scanning Kerr microscopy (TRSKM) was used to image the out of plane magnetization induced by $I_{DC}$ and $I_{RF}$ within injection-locked nano-contact (NC) SHNO devices\cite{Spicer2018a}. Each NC-SHNO was formed on a Py/Pt bilayer disk, with $I_{DC}$ concentrated within a small central region.  While a non-linear bullet mode was observed at the device centre when $I_{DC}$ exceeded a threshold value,  $I_{RF}$ was observed to stimulate a delocalised and spatially inhomogeneous response.  Here TRSKM is used to determine the spatial variation of the torques generated by the $I_{RF}$ injected into a NC-SHNO.  The spatial variation of both the STT and Oersted torques is found to diverge strongly from that expected for the DC current distribution, suggesting that the reactance of the device geometry strongly modifies the RF current distribution.

\begin{figure}[h]
\includegraphics[width=8.5cm]{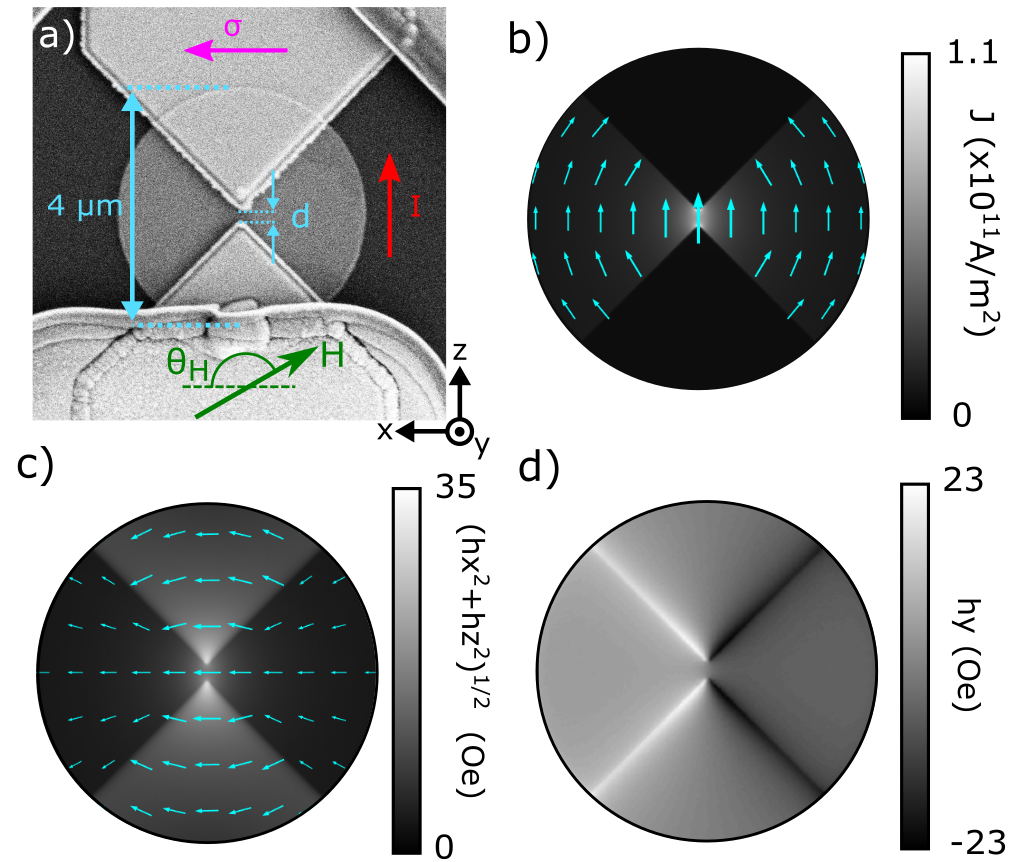}
\caption{(a) SEM image of a NC-SHNO showing electrode separation $d$, directions of the DC charge current $I_{DC}$,  average polarization \textbf{$\hat{\sigma}$} of the spin current injected into the Py layer, and field $\textbf{H}$ orientated at angle $\theta_H$ relative to the $x$ axis. COMSOL simulations for $I_{DC}$ = 4 mA and $d$= 200 nm are shown for (b) current density $\textbf{J}$ within the Pt layer, (c)  in-plane Oersted field within the Py layer that has magnitude $\sqrt{h_x^2+h_z^2}$, and (d),  out of plane Oersted field $h_y$ within the Py layer. The arrows and grayscale indicate the direction and normalized magnitude respectively.}
\label{Fig:1}
\end{figure}

NC-SHNO devices, shown in Figure 1, were fabricated on sapphire substrates by sputtering and electron-beam lithography  \cite{Durrenfeld2015}. Triangular needle-shaped Au(150 nm) NCs with tip separation of $d$ = 140 - 240 nm were defined on a 4 $\mu$m diameter Py(5 nm)/Pt(6 nm) bi-layer disk. The $I_{DC}$ from the NCs is concentrated within a small region of the Pt layer between the tips, and generates a spin current that flows into the Py layer beneath. The injected spin polarization lies parallel to the +ve $x$ direction along a horizontal line through the middle of the disk,  \cite{Emori2013} and exerts a STT on the Py magnetization. The charge current also generates an Oersted field with in and out of plane components.  The distributions of $I_{DC}$ and the Oersted field plotted in Figure \ref{Fig:1} were calculated using COMSOL  \cite{COMSOL}.  The STT amplitude has similar spatial distribution to the charge current within the Pt, while the Oersted field depends upon the current distribution in both the Pt and Au layers.

Conventional STT-FMR measurements were made by applying audio frequency modulation to the $I_{RF}$ injected through the capacitative arm of a bias-tee, while $V_{mix}$ was measured through the inductive arm using a lock-in amplifier. The out-of-plane component of the dynamic magnetization was detected directly by means of TRSKM that has been described elsewhere \cite{Keatley2017}.  

Both the dynamic magnetization and $V_{mix}$ can be calculated for a macrospin. The equation of motion for the magnetization driven by an external field and STT is  

\begin{multline}
\frac{d\hat{\textbf{m}}}{dt}=-|\gamma|(\hat{\textbf{m}}\times\textbf{H}_{eff})+\alpha\hat{\textbf{m}}\times\frac{d\hat{\textbf{m}}}{dt}
\\
-|\gamma|A\hat{\textbf{m}}\times(\hat{\textbf{m}}\times\hat{\mathbf{\sigma}})+|\gamma| B(\hat{\textbf{m}}\times\hat{\mathbf{\sigma}}),
\label{EQ:LLG}
\end{multline}

where $\textbf{\^{m}}$ is the normalized magnetization vector, $\gamma$ is the gyromagnetic ratio, $\alpha$ is the Gilbert damping constant, $\hat{\mathbf{\sigma}}$ is the injected spin polarisation, and $A$ and $B$ are the amplitudes of the ``in-plane" or ``anti-damping", and ``out-of-plane" or ``field-like" STT respectively. A large in-plane torque is expected due to the SHE. However, since the Py is relatively thick and the current is shunted through the Pt, negligible torque is expected due to the Rashba effect. $\textbf{H}_{eff}$ is the total effective field acting upon the magnetization, which may be written as $\textbf{H}_{ext}+H_d(\textbf{\^{y}}\cdot\textbf{\^{m}})\textbf{\^{y}}+\textbf{h}$, where $\textbf{H}_{ext}$ is the static applied field, $H_d$ is the out of plane demagnetizing field, and $\textbf{h}$ is the local Oersted field generated by the RF current. Other anisotropy fields are expected to be small and have been neglected.


Equation (1) can be linearized for small amplitude precession, with the out of plane magnetization written as $m_y(\varphi)=Re(ae^{i\varphi})$ where $\varphi$ is the phase of the RF current and $a$ is the complex amplitude. $V_{mix}$ and the real and imaginary parts of $a$ have the forms


\begin{widetext}
\begin{equation}
V_{mix} = I_{RF}\\\Delta R sin\theta_H cos\theta_H
\\
\frac{|\gamma'|^2 H_{\perp}(f_0^2-f^2)(H_{ext}+H_d)
+|\gamma'|f^2\Delta(\alpha H_{\perp}+H_{||})}{(f_0^2-f^2)^2+f^2\Delta^2},
\label{EQ:Vmix}
\end{equation}

\begin{equation}
Re(a) = 
\frac{-|\gamma'|^2H_{ext}H_{||}(f_0^2-f^2)+f^2\Delta|\gamma'|(H_{\perp}-\alpha H_{||})
}{(f_0^2-f^2)^2+f^2\Delta^2},
\label{EQ:Re(a)}
\end{equation}

\begin{equation}
Im(a) = 
\frac{f|\gamma'|(H_{\perp}-\alpha H_{||})(f_0^2-f^2)+ f\Delta|\gamma'|^2H_{ext} H_{||}
}{(f_0^2-f^2)^2+f^2\Delta^2},
\label{EQ:Im(a)}
\end{equation}

where

\begin{equation}
H_{\perp}=\left(sin\theta_H h_x+B sin\theta_H\right), H_{||}=\left(\frac{sin\theta_H}{|sin\theta_H|}h_y-A sin\theta_H\right),
\label{EQ:PeraPerp}
\end{equation}

\end{widetext}

$\gamma'=\gamma/2\pi$, $f$ and $I_{RF}$ are the frequency and amplitude of the RF current, $\theta_H$ is the angle between $\textbf{H}$ and \textbf{$\hat{\sigma}$}, $H_{\perp}$ and $H_{||}$ represent effective fields, where the subscripts indicate the direction in which the associated torque acts, and $A$ and $B$ are defined in equation (\ref{EQ:LLG}). Finally, the linewidth $\Delta=|\gamma'|\alpha(2H_{ext}+H_d)$, and the FMR frequency $f_0=|\gamma'|\sqrt{H_{ext}[H_{ext}+H_d]}$. $\Delta R$ = 0.03 $\Omega$ \cite{Durrenfelda} is the change in electrical resistance when the magnetisation is rotated from orthogonal to parallel to the current.   

The above expressions yield $m_y$ at different positions within the NC-SHNO if the observed dynamical magnetization is a response to local torques.  This is a reasonable assumption when spin waves excited due to spatially varying STT and Oersted torques are similar in frequency. Dispersion due to dipolar interactions decreases with film thickness. For the 5 nm Py film, the frequency splitting, of the uniform mode and a spin wave with wavelength equal to the diameter of the disk, is no more than $20\%$ and lies within the measured linewidth.  

The stroboscopic nature of TRSKM requires that measurements are made at an RF frequency that is a multiple of the laser repetition rate as $H_{ext}$ and hence $f_0$ are varied. From equations (\ref{EQ:Re(a)}) and (\ref{EQ:Im(a)}), $Re(a)$ and $Im(a)$ are seen to contain a minimum in the denominator at the resonance field, and terms in the numerator that are either slowly varying or antisymmetric (due to the factor $f_{0}^2-f^2$) about the resonance field.  Hence both expressions consist of parts that are symmetric and antisymmetric about the resonance field. The microwave phase $\varphi$ may be chosen in the experiment, so that $m_y$ is a weighted sum of $Re(a)$ and $Im(a)$, and hence a sum of symmetric and antisymmetric terms. TRSKM measures the polar Kerr rotation  $QM\hat{m}_{y}$ where the constant $Q$ is of order 0.1 mdeg cm$^3$ emu$^{-1}$. If the value of $Q$ is known then, by recording the dependence of $m_y$ upon $H_{ext}$ for a number of values of $\varphi$, and fitting $m_y$ to equations (\ref{EQ:Re(a)}), (\ref{EQ:Im(a)}) and (\ref{EQ:PeraPerp}), the values $A$ and $B$ can be determined at each position within the sample.

\begin{figure}
\includegraphics[width=8.5cm]{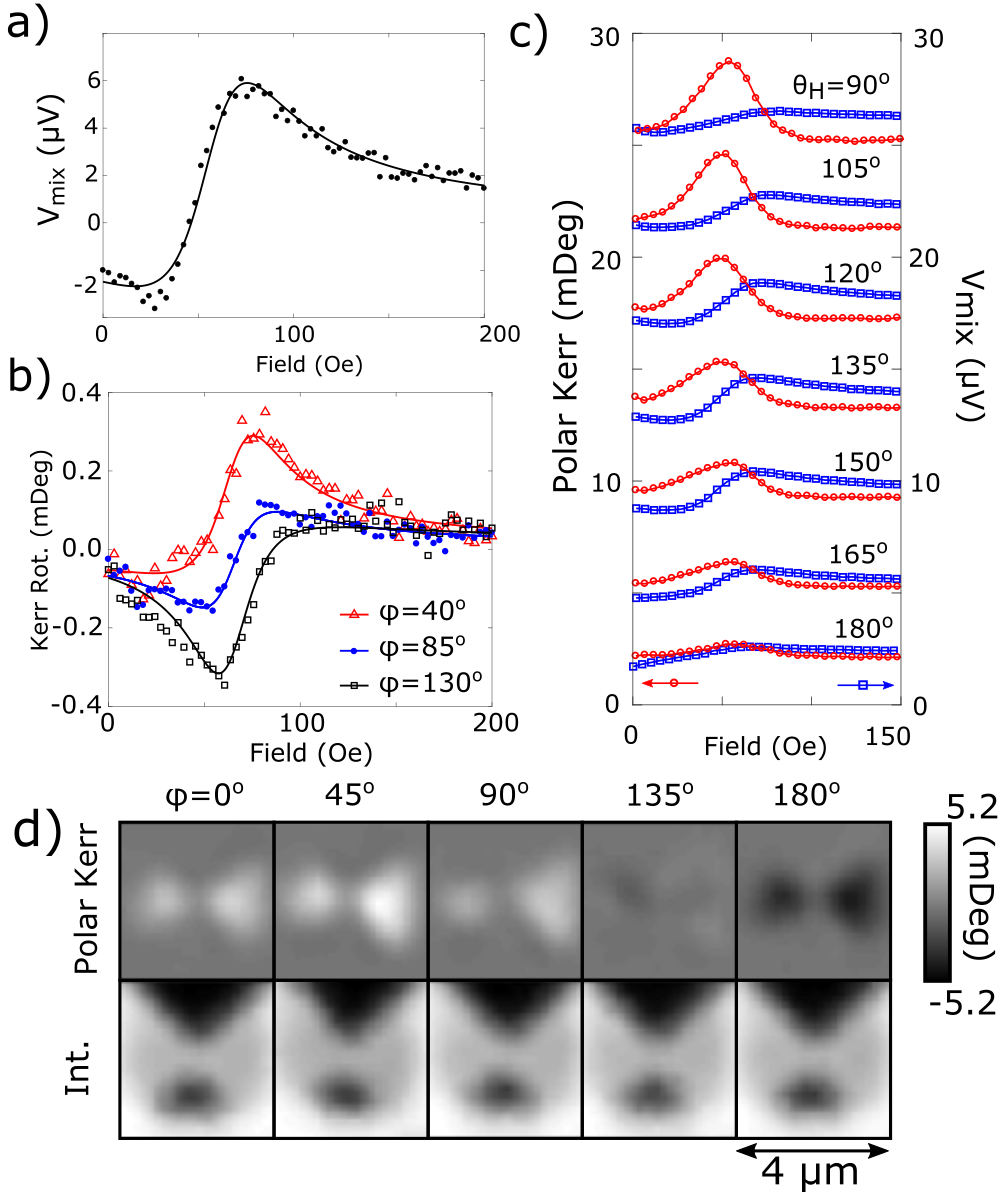}
\caption{(a) Dependence of mixing voltage $V_{mix}$ upon field $H_{ext}$ applied at $\theta_H=150^\circ$. The continuous curve is fit to equation \ref{EQ:Vmix}. (b) Polar Kerr rotation curves recorded with the optical spot between the tips of the NC for three values of the phase $\varphi$. (c) Polar Kerr (closed circles) and $V_{mix}$ (open squares) resonance curves recorded with the optical probe $1 \mu$m to the right of centre, for different values of $\theta_H$. (d) Polar Kerr rotation and intensity images for $\theta_H=90^\circ$ and $H_{ext} = 75 $Oe for $\varphi$ values in the range 0 to $180^\circ$. $I_{RF}$ had frequency of 2 GHz and amplitude of 2.8, 1.3, 3.2 and 4.0 mA in (a), (b), (c) and (d) respectively. The NC separation $d=$ 240 nm in (a) and (b), and 140 nm in (c) and (d)}
\label{Fig:combined}
\end{figure}

Conventional STT-FMR was first performed to obtain $V_{mix}$, as shown in Figure \ref{Fig:combined}(a). The optical probe was then positioned between the NC tips and the polar Kerr signal recorded at three values of RF phase as a function of field, as shown in Figure \ref{Fig:combined}(b). Both optical and electrical resonance curves are a superposition of components that are either symmetric or antisymmetric about the resonance field. The  $V_{mix}$ data, which does not depend upon $\varphi$, was fitted to equations (\ref{EQ:Vmix}) and (\ref{EQ:PeraPerp}), yielding values of $H_{\perp} = -6.1 \pm 0.8$ Oe and $H_{||} = 29.7 \pm 3$ Oe, while values of $|\gamma'|=2.94$ MHz/Oe, $\alpha=0.04$ and $H_D=8000$ Oe were found to best describe both the $V_{mix}$ and optical data within the present study \cite{SupMat}. Measurements made at different $I_{RF}$ values confirmed that the dependence of the extracted effective fields upon $I_{RF}$ was roughly linear for the range of values studied here \cite{SupMat}. The large value of $\alpha$ has been attributed to spin pumping effects \cite{Liu2013}. Since the average value of $h_y$ is small due to the NC symmetry, the large value of $H_{||}$ results from the anti-damping torque. From equation (\ref{EQ:PeraPerp}), if $\theta_H = 150^\circ$, $h_x$ has value of $\sim$12.1 Oe, similar to the 10.2 Oe calculated by COMSOL at the centre of the disk for a DC current. 

Due to the symmetry of the device, one may assume \cite{SupMat} that the ratio $H_{\perp}/H_{||}$ determined from optical measurements in Figure \ref{Fig:combined}(b) should be the same as that determined by fitting $V_{mix}$. Fixing this ratio and fitting the optical resonance curves yields values for $\varphi$, that have been used to label each curve. The three phase values were found to be offset set by the same amount from the values set on the microwave synthesizer \cite{SupMat}, justifying the assumed value of  $H_{\perp}/H_{||}$.  The fitting also yields $Q$, but this is less reliable because the areas sampled by the electrical and optical measurements are different, as will be discussed further.

The dependence of  optical and electrical signal strength upon $\theta_H$ is shown in Figure \ref{Fig:combined}(c).  Maximum optical signal amplitude is observed for $\theta_H=90^\circ$, due to the sin$\theta_H$ factor in equation (\ref{EQ:PeraPerp}). In contrast $V_{mix}$ vanishes for $\theta_H= 90^\circ$ and $180^\circ$, and so is  insensitive to the dynamics when $\theta_H=90^\circ$. Finally, polar Kerr images are plotted in Figure \ref{Fig:combined}(d) for different values of $\varphi$. Due to the symmetry of the current distribution about a vertical line though the centre of the device, $A$, $h_x$ and hence $H_{\perp}$  are expected to be symmetric about this centre line.  On the other hand $h_y$ is antisymmetric so that $H_{||}$ has mixed symmetry. If terms in $\alpha$ are neglected in equations (\ref{EQ:Re(a)}) and (\ref{EQ:Im(a)}), then at resonance, when $f = f_0$, $Re(a)$ is symmetric about the centre line, while $Im(a)$ has mixed symmetry.  Therefore, the most symmetric image should occur for $\varphi = 0^\circ$. The most striking feature of the images is the minimum at the NC tips, which is unexpected from the calculation of $I_{DC}$ in Figure \ref{Fig:1}. 

\begin{figure}
\includegraphics[width=8.5cm]{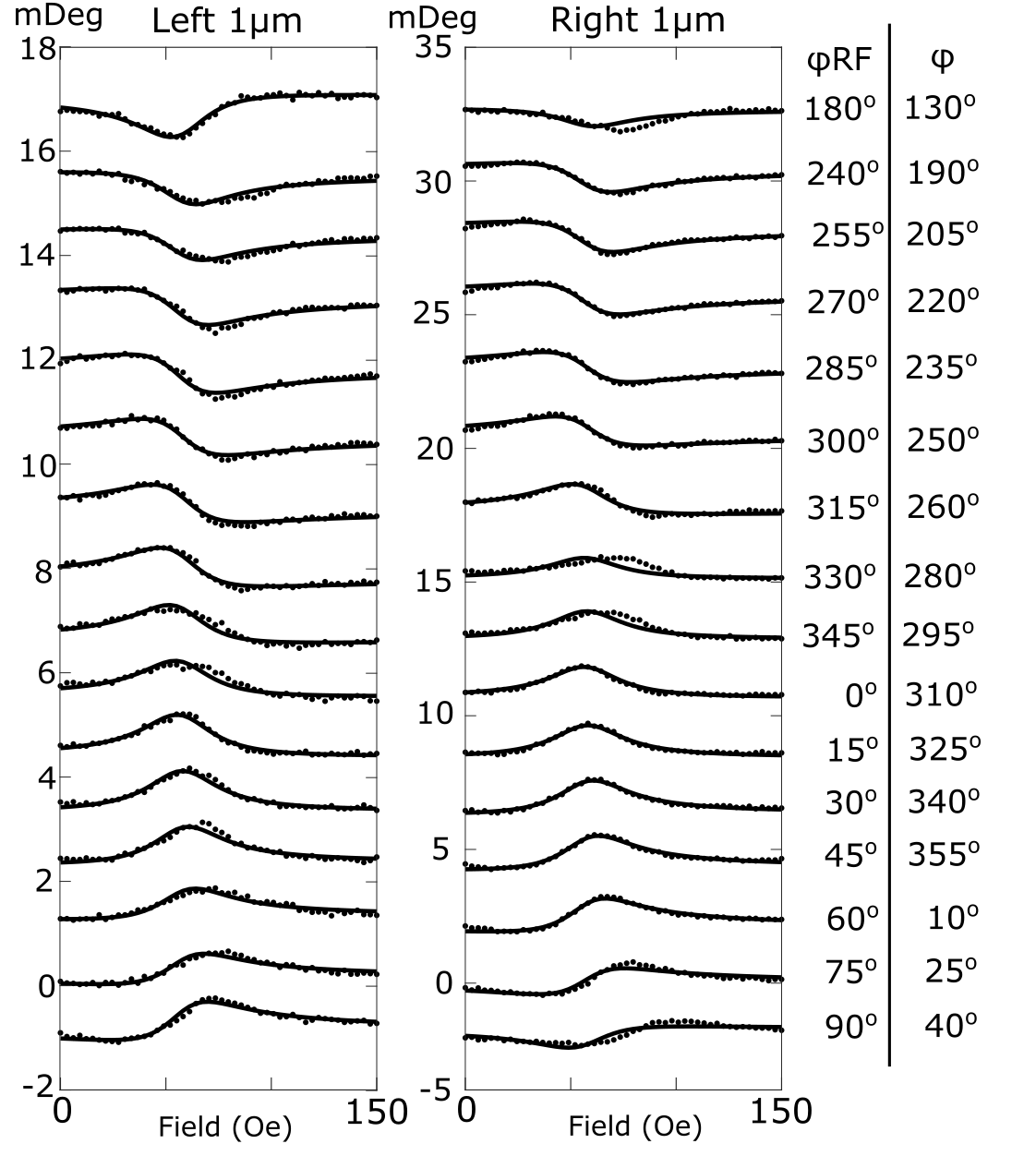}
\caption{Optically-detected resonance curves acquired at points 1 $\mu$m to the left and right of the centre of the device, for selected values of RF phase, with $\theta_H$ = 90$^\circ$, $f$ = 2 GHz, $I_{RF}$ = 1.3 mA, and $d$ = 240 nm. $\varphi_{RF}$ is the phase value set at the microwave synthesizer, while $\varphi$ is the phase of $I_{RF}$ at the device. The solid symbols represent the measured Kerr rotation values, while the continuous curves are fits that are described within the main text.}
\label{Fig:PhaseVar}
\end{figure}

\begin{table}
\centering
\begin{tabular}{|c|c|c|}
\hline 
 & Mean $H_\perp$ (Oe) & Mean $H_{||}$ (Oe) \\ 
\hline 
Left & 0.8935 $\pm$ 0.0997 & -5.3400 $\pm$ 0.8643 \\ 
\hline 
Right & 0.9141 $\pm$ 0.4807 & -13.4049 $\pm$ 1.7338 \\ 
\hline 
\end{tabular} 
\caption{Mean and standard deviation of the effective field parameters extracted from the fits shown in Figure \ref{Fig:PhaseVar}}.
\label{tab:PhaseParams}
\end{table}

Optically-detected resonance curves were acquired with the probe spot placed 1 $\mu$m to either side of the centre of the device to demonstrate the spatial and phase sensitivity of the measurement technique. Figure \ref{Fig:PhaseVar} shows the resonance curves with fits that assume, as before, that the phase  of $I_{RF}$ at the device $\varphi$ is offset by 50$^\circ$ from the phase set on the microwave synthesizer $\varphi_{RF}$. The different fitted curves assume the same set of experimental and material parameter values apart from $H_\perp$ and $H_{||}$, which were allowed to vary for each curve. The fits generally describe the data very well, with a few exceptions where either phase wander of the microwave synthesizer or drift in the position of the focused laser spot may have led to distortion of the experimental data. The mean and standard deviation of the extracted effective field values are plotted in table \ref{tab:PhaseParams}. The extracted values for $H_\perp$ are seen to be reasonably consistent for all values of the phase and on both sides of the disk, justifying the method used to determine the phase offset. In contrast $H_{||}$ is significantly different between the two sides of the disk, as will be discussed further below.

\begin{figure}
\includegraphics[width=8.5cm]{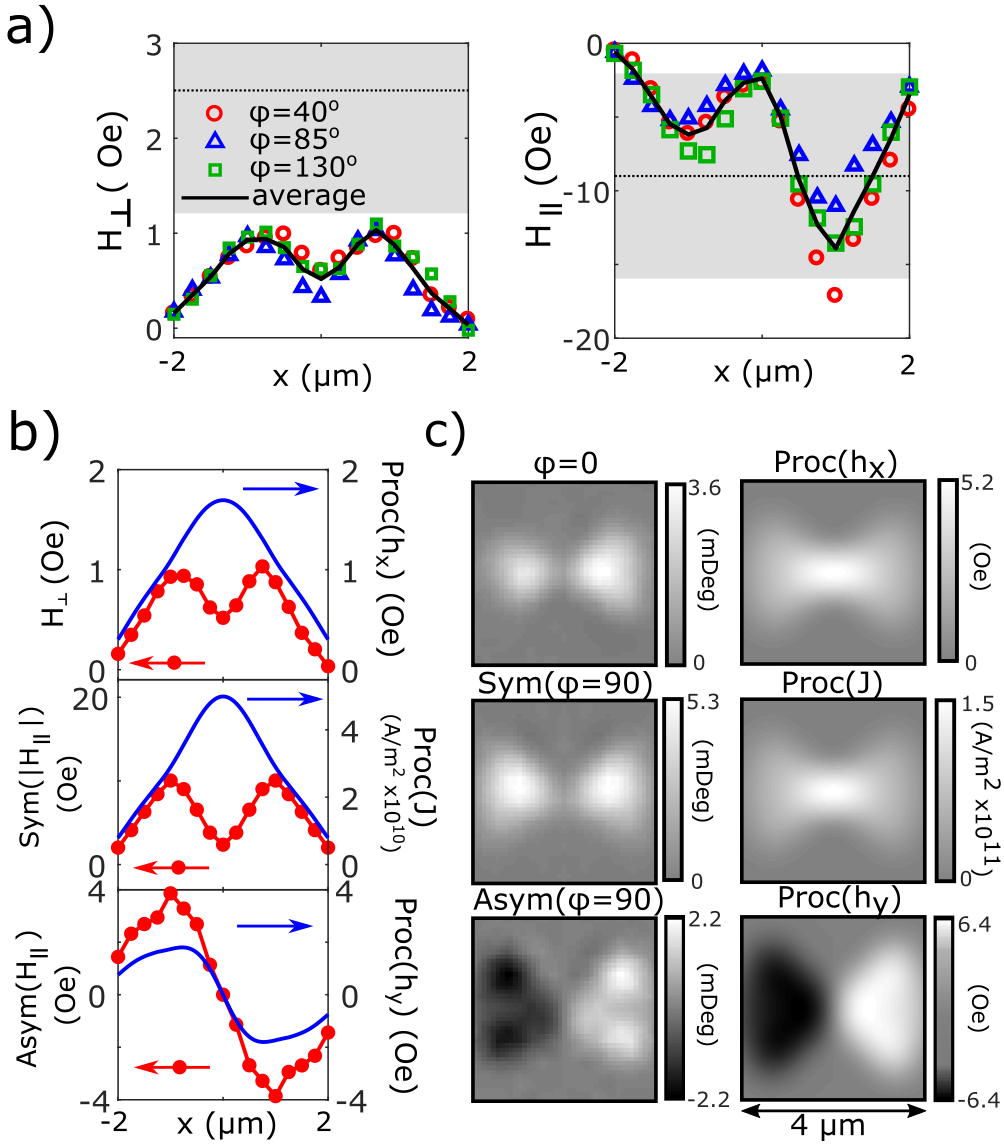}
\caption{(a) Spatial variation of $H_{\perp}$ and $H_{||}$ along the horizontal line passing through the middle of the disk. Values obtained from electrical STT-FMR are shown by the dashed line with the error indicated by the gray shading. The injected $I_{RF}$ had frequency $f$ = 2 GHz  and amplitude of 1.3 mA, while the static field was applied at $\theta_H$ = 90$^\circ$. (b) $H_{\perp}$ and the antisymmetric/symmetric parts of $H_{||}$ are plotted together with the current and Oersted field from Figure 1. (c) Polar Kerr image acquired for $\varphi=0^\circ$, $I_{RF}$ = 4 mA and $H_{ext}$ = 75 Oe is presented next to the calculated distribution of $h_x$. Symmetric and antisymmetric parts of the polar Kerr data acquired at $\varphi=90^\circ$, are presented next to the vertical component of the current density $J_z$ and the out of plane Oersted field $h_y$. The calculated images were convolved with a Gaussian profile of 870 nm half maximum diameter. The NC separation $d$ was 240 nm in (a) and (b), and 140 nm in (c).}
\label{fig: distvar}
\end{figure}

To further explore the spatial symmetry of the magnetic response, and hence the underlying torques, the field dependence of the polar Kerr rotation was measured at different points on a horizontal line through the middle of the disk, for values of $\varphi = 40^\circ$, 85$^\circ$ and 130$^\circ$. The extracted values of $H_{\perp}$ and $H_{||}$ are plotted in Figure \ref{fig: distvar}(a). The field values obtained for the three $\varphi$ values are in good agreement confirming that the absolute phase has been determined correctly.

Since negligible out of plane STT is expected, $H_{\perp}$ should be proportional to $h_x$ and spatially symmetric. However, $H_{||}$ should contain both  symmetric and antisymmetric components, (denoted as $Sym(|H_{||}|)$ and $Asym(H_{||})$) due to the in plane STT and $h_y$ respectively. These components can be separated by calculating the mirror image (reflection about $x=0$) of the $H_{||}$ data, calculating the sum and difference of the original data with its mirror image, and then dividing both by a factor of two. Both components are plotted in Figure \ref{fig: distvar}(b), together with convolutions of the Oersted field and current distribution of  Figure \ref{Fig:1} with a Gaussian function of 870 nm half maximum width (denoted Proc($h_{x,y}$) and Proc($J$)\cite{SupMat}), to account for the size of the optical spot. In extracting field quantities from the data, a value of $Q$ = 0.3 mdeg cm${^3}$ emu$^{-1}$ was assumed so that the Oersted fields towards the edge of the disk were in agreement with those from Figure \ref{Fig:1}.  However, the experimental $H_{\perp}$ and $Sym(|H_{||}|)$ curves exhibit a minimum at $x = 0$ $\mu$m, in disagreement with the calculated curves.

As explained above, for $\varphi = 0^\circ$ and $90^\circ$, time resolved images acquired at resonance reveal the spatial variation of $H_{\perp}$ and $H_{||}$ respectively. A similar procedure to that applied to the line scan in Figure \ref{fig: distvar}(b) was used to extract the symmetric and antisymmetric parts of the image acquired at $\varphi=90^\circ$.  The resulting images of $H_{\perp}$, $Sym(|H_{||}|)$ and $Asym(H_{||})$ are plotted next to calculated images of $h_x$, $J$, and $h_y$ in Figure \ref{fig: distvar}(c). Each calculated distribution has been convolved with a 2D Gaussian function of 870 nm half maximum diameter. The form of the $Asym(H_{||})$ and Proc($h_y$) images are in reasonable agreement.  However the Proc($h_x$) and Proc($J$) images possess a maximum at the centre of the disk, whereas the $H_{\perp}$ and $Sym(|H_{||}|)$ images exhibit a minimum.  The convolution with the spot profile accounts for the NCs partly obscuring the underlying Py/Pt bilayer, so that the minimum corresponds to a reduction of the in-plane STT and the torque due to the in-plane Oersted field. 
The observed minimum initially seems at odds with observations of a self-localised bullet mode at the centre of this \cite{Spicer2018a,Durrenfelda} and similar devices \cite{Demidov2012},  but may be explained if the spatial distribution of $I_{RF}$ is different to that of $I_{DC}$ used to excite the bullet mode.

\begin{figure}[h!]
\includegraphics[width=8.5cm]{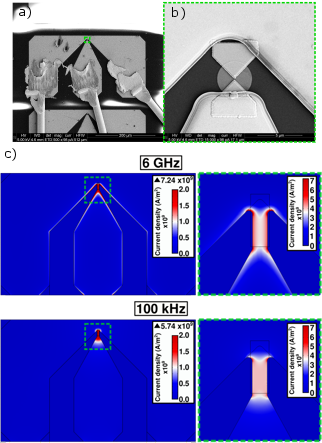}
\caption{(a), (b) Scanning electron microscope images of NC-SHNO devices.  In (a) wire bonds have been made to the ground-signal-ground pads of a coplanar waveguide structure. The signal line tapers to where the NC-SHNO device is located.  The green dashed box indicates the region that is shown at greater magnification in (b) where a device is located between the end of the tapered signal line and the point at which the two ground planes meet. (c) Simulated distribution of $I_{RF}$ within a coplanar waveguide (CPW) structure with similar lateral dimensions to that of the NC-SHNO device structure at  frequencies of 100 kHz and 6 GHz. The CPW is 1 um thick, has a gold-like conductivity of 45.6 $\times$10$^6$ S/m, and is surrounded by a large air region with volume of 450 $\times$ 450 $\times$ 400 $\mu$m$^3$. It is excited by a coaxial lumped port in a ground-signal-ground configuration producing an input current of 16 mA. A narrow constriction of 4 $\mu$m width is located at the end of the tapered signal line.  The plotted quantity is the magnitude of the current density in the plane passing through the middle of the gold layer. For clarity the range of the colour scale has been adjusted in the main images, with maximum values indicated above the relevant legends. The colour scales of the 100 kHz plots have been made similar to those of the 6 GHz plots to aid comparison.}
\label{Fig:RFModel}
\end{figure}

{The NC-SHNO devices were located in a constriction of the signal line of a shorted coplanar waveguide (CPW) as shown in Figure} \ref{Fig:RFModel} (a) and (b).  It is well established that the distributed reactance of the CPW gives rise to spreading of current from the centre of the signal line and hence to current crowding at its edges \cite{Chia-NanChang1990,Carlsson1997,Carlsson1999}. The distribution of $I_{RF}$ within the NC-SHNO results from competition between the confining effect of the NCs and current spreading within the larger scale device structure that has the form of a short CPW.  Modelling the spatial distribution of $I_{RF}$ within a NC-SHNO using the COMSOL package is very difficult due to the need to mesh a structure with feature sizes that vary from the order of 1 nm to 100 $\mu$m.  Instead a simulation was performed for a structure that had lateral dimenions similar to that of the NC-SHNO device structure, but which was formed from a layer of gold with uniform thickness of 1 $\mu$m, as shown in Figure \ref{Fig:RFModel} (c). The circular mesa and NCs were replaced by a constriction of the signal line of 4 $\mu$m width.  Simulations for frequencies of 100 kHz and 6 GHz are shown.  While the current is uniformly distributed across the width of the constriction at 100 kHz, spreading towards the edge of the signal line is observed at 6 GHz, establishing the feasibility of the current spreading mechanism within the NC-SHNO device.  The reduction of the torque due to a reduction of $I_{RF}$ at the centre may explain why the  frequency range for injection-locking of auto-oscillations observed in NC-SHNOs is reduced relative to that for nano-constriction SHNOs. \cite{Demidov2014}

In summary, it has been shown that TRSKM can be used to probe the local FMR driven by a combination of STT and Oersted field torques, and comparison has been made with a simple theory.  By directly probing the local magnetization, this technique can be applied to magnetic materials or experimental configurations that exhibit weak MR response. Furthermore the phase and spatial symmetry of the different torques allows them to be separated and mapped. It appears that the spatial distributions of $I_{DC}$ and $I_{RF}$ are significantly different, perhaps due to the reactance of the device that leads to spreading of $I_{RF}$.  The torques due to the RF current exhibit a minimum at the centre of the device, which suggests that the RF current may act to destabilise the position of the self-localised bullet mode, and may explain why the NC-SHNO exhibits a reduced locking range relative to other STOs. The present study emphasises the need to characterise both the DC and RF current distributions and the torques that they generate.

\begin{acknowledgments}
We acknowledge financial support from the Engineering and Physical Sciences Research Council (EPSRC) of the United Kingdom, via the EPSRC Centre for Doctoral Training in Metamaterials (Grant No. EP/L015331/1), and grants EP/I038470/1 and EPSRC EP/P008550/1. All data created during this research are openly available from the University of Exeter’s institutional repository at https://doi.org/10.24378/exe.1003.
\end{acknowledgments}

\bibliography{library_Adjust}  

\end{document}